\documentclass{osa-article}

\journal{oe}

\usepackage{caption}
\usepackage{subcaption}
\usepackage{booktabs}
\usepackage{lineno}
\usepackage{amsmath}

\newcommand{\tripletDJ}[1]{$^3D_{#1}\,$}

\newcommand{\raI}{$^{226}\text{Ra}$}
\newcommand{\raII}{$^{225}\text{Ra}$}

\newcommand{\singletSzero}{$^{1}S_{0}$\,}
\newcommand{\tripletPone}{$^{3}P_{1}$\,}
\newcommand{\singletPone}{$^{1}P_{1}$\,}

\newcommand{\radI}{$^{226}\text{Ra}$\,}
\newcommand{\radII}{$^{225}\text{Ra}$\,}

\begin{document}

\title{Implementing an electronic sideband offset lock for {isotope shift} spectroscopy in radium}

\author{T. Rabga,\authormark{1,2,3,*} K. G. Bailey,\authormark{1} M. Bishof,\authormark{1,\textdagger} D. W. Booth,\authormark{1} M. R. Dietrich,\authormark{1} J. P. Greene,\authormark{1} P. Mueller,\authormark{1} T. P. O'Connor,\authormark{1} and J. T. Singh\authormark{2}}

\address{\authormark{1}Physics Division, Argonne National Laboratory, Argonne, Illinois 60439, USA\\
\authormark{2}National Superconducting Cyclotron Laboratory and Department of Physics and Astronomy, Michigan State University, East Lansing, Michigan 48824, USA\\
\authormark{3}Currently with the Center for Correlated Electron Systems, Institute for Basic Science (IBS) and Department of Physics and Astronomy, Seoul National University (SNU),
Seoul 151-742, Republic of Korea}

\email{\authormark{*}trabga@snu.ac.kr} 
\email{\authormark{\textdagger}bishof@anl.gov}



\begin{abstract}

We demonstrate laser frequency stabilization with at least 6 GHz of offset tunability using an in-phase/quadrature (IQ) modulator to generate electronic sidebands (ESB) on a titanium sapphire laser at 714 nm and we apply this technique to perform {isotope shift} spectroscopy of \raI\, and \raII. By locking the laser to a single resonance of a high finesse optical cavity and adjusting the lock offset, we determine the frequency difference between the magneto-optical trap (MOT) transitions in the two isotopes to be $2630.0\pm0.3$ MHz, a factor of 29 more precise than the previously available data. Using the known value of the hyperfine splitting of the \tripletPone level, we calculate the isotope shift for the \singletSzero to \tripletPone transition to be $2267.0\pm2.2$ MHz, which is a factor of 8 more precise than the best available value. Our technique could be applied to countless other atomic systems to provide unprecedented precision in isotope shift spectroscopy and other relative frequency comparisons.  
\end{abstract}

\section{Motivation and background}

Laser frequency stabilization techniques are ubiquitous in applications such as precision spectroscopy~\cite{Hansch99}, laser cooling and trapping of atoms~\cite{Phillips98} and molecules~\cite{Jin2012}, and quantum information science~\cite{monroe10}. Laser frequency stabilization is often achieved via comparison to a stable frequency reference.  The most common frequency references include optical cavities and atomic or molecular transitions, but any system with a stable, measurable, and selective response to laser frequency can be used.   This response signal, can then be used as an "error signal," which tracks the laser's frequency deviations from the stable reference and can be "fed back" to parameters that control the laser frequency to cancel these deviations.  In this work, we use the reflection signal from a high-finesse optical cavity near a cavity resonance as our frequency reference.

In a side-of-peak locking scheme, one locks the laser frequency to the side of an optical cavity resonance - using the slope of the cavity resonance as the error signal. Although this allows some tunability in the frequency of the laser, by selecting different positions on the resonance, it couples any laser intensity fluctuations to its frequency instability. A preferred and an improved method uses the Pound-Drever-Hall (PDH) locking scheme \cite{Drever83,Black01}. This involves the phase modulation of the laser beam incident on a optical cavity reference. The reflected signal from the cavity is collected on to a photo detector and demodulated to generate the error signal. This method overcomes the sensitivity to intensity fluctuations from the side-of-peak scheme at the price of tunability, since insensitivity to laser intensity noise is optimal only at the resonance peak.  

However, it is still often desirable to have a tunable laser frequency lock while maintaining frequency stability. There are a variety of ways to achieve this goal. Using an acousto-optic modulator (AOM), one can achieve several hundreds of MHz of frequency tunability\cite{Bondu97}. Alternatively an offset phase lock to another frequency stabilized laser can achieve tunable frequency locks over a larger tuning range\cite{Ye99}, but requires an additional laser. The electronic side-band (ESB) offset lock, a simple extension to the PDH lock, allows laser frequency stabilization to a fixed frequency reference with a broadly tunable offset frequency \cite{Thorpe08,Kohlhaas:12,Milani:17}. Offset frequencies up to 4 GHz have been achieved by combining the ESB offset locking technique with a high-bandwidth, fiber-coupled electro-optical modulator (EOM)\cite{Bai17}. Here, we describe the methods used to implement an ESB offset lock for laser frequency stabilization with an offset frequency that is tunable between 200 MHz and 6 GHz.  In contrast to previous work, we use an in-phase/quadrature (IQ) modulator (Analog Devices, LTC5588-1) to generate the laser modulation signal from inexpensive digital signal generators.  This eliminates the need for expensive microwave signal generators capable of carrier phase modulation at several MHz. Moreover, our approach can be adapted to a greater number of applications owing to the broad availability of low cost synthesizers and IQ modulators across the entire RF and Microwave spectrum. We leverage our ESB offset lock to perform laser spectroscopy of radium isotopes and compare the frequencies of different atomic transitions to the same optical cavity resonance. We determine the frequency difference between isotopes with precision and accuracy limited only by the cavity resonance, lock implementation, and fundamental atomic properties. 

\begin{figure}[ht]
    \centering
    \includegraphics[width=0.6\textwidth]{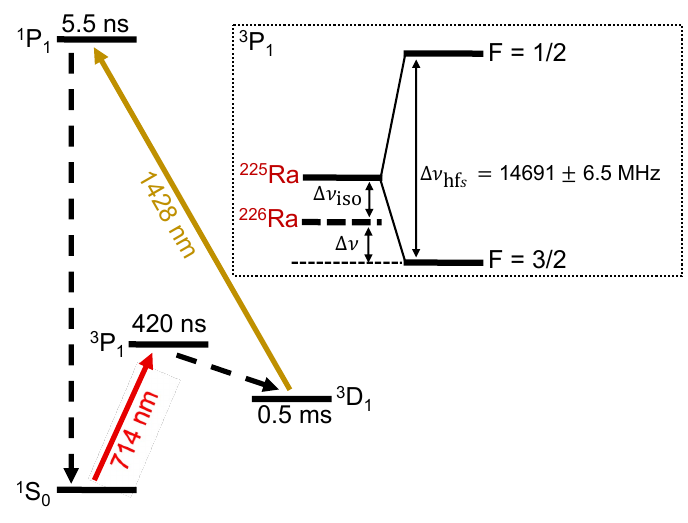}
    \caption{The atom slowing and cooling scheme used to trap \radI and \radII in a MOT. The \singletSzero $\rightarrow$ \tripletPone transition requires a single repump laser at 1428 nm to excite the \tripletDJ{1} $\rightarrow$ \singletPone transition and close the cooling cycle. Inset: The frequency difference $\Delta \nu$ between the MOT transitions for the two isotopes is shown. Also shown are the isotope shift for the transition $\Delta \nu_{\text{iso}}$ and the hyperfine splitting $\Delta\nu_{\text{hfs}}$ in \raII.}
    \label{fig:redslower}
\end{figure}

Spectroscopy of radium isotopes is of extreme importance due to their unique atomic and nuclear properties that make them suitable for electric dipole moment (EDM) searches. A non-zero EDM in a non-degenerate system violates time-reversal (T) symmetry and consequently charge-parity (CP) symmetry~\cite{chupp19}. The octupole deformation and nearly degenerate nuclear parity doublet in radium make it an ideal candidate for probing CP violations due to the atomic nucleus \cite{engel05, singh15, Dzuba2009, auerbach96}. The radium EDM experiment at the Argonne National Laboratory uses \raII\,($I=1/2$, $\tau_{1/2} = 14.9$ day) (where $I$ is the nuclear spin) to set the best limit on the size of the EDM in \raII \cite{mike16}. Its 14.9-day half-life and low vapor pressure make \raII\, a challenging system for an EDM experiment. Due to its greater abundance and longer half-life, \raI\, ($I=0$, $\tau_{1/2} = 1600$ yr) is used to optimize certain parts of the EDM experimental apparatus, such as the atom cooling and trapping setup. Therefore it is crucial to be able to quickly and reliably tune laser frequencies over several GHz in order to laser cool and trap both isotopes. The laser frequency stabilization technique described here provides a convenient and robust method for tuning the laser frequency to the relevant atomic transitions used for cooling and trapping these two isotopes during a single experiment.

The relevant energy levels for slowing and trapping of a radium magneto-optical trap (MOT) are shown in Figure~\ref{fig:redslower}~\cite{mike16}. We perform an improved measurement of the frequency difference $\Delta \nu$ between the MOT transitions for \radI (\singletSzero $\rightarrow$ \tripletPone) and \radII (\singletSzero [F=1/2] $\rightarrow$ \tripletPone [F=3/2]). From the measured $\Delta \nu$ and the known value for the hyperfine splitting $\Delta\nu_{\text{hfs}}$ between the $F=3/2$ and $F=1/2$ levels of \tripletPone, we extract the isotope shift $\Delta\nu_{\text{iso}}$ between \raI\, and \raII\, for the \singletSzero $\rightarrow$ \tripletPone transition.

\section{Implementing an ESB offset lock using an IQ modulator}

\begin{figure}[ht]
    \centering
    \includegraphics[width= 0.7\textwidth]{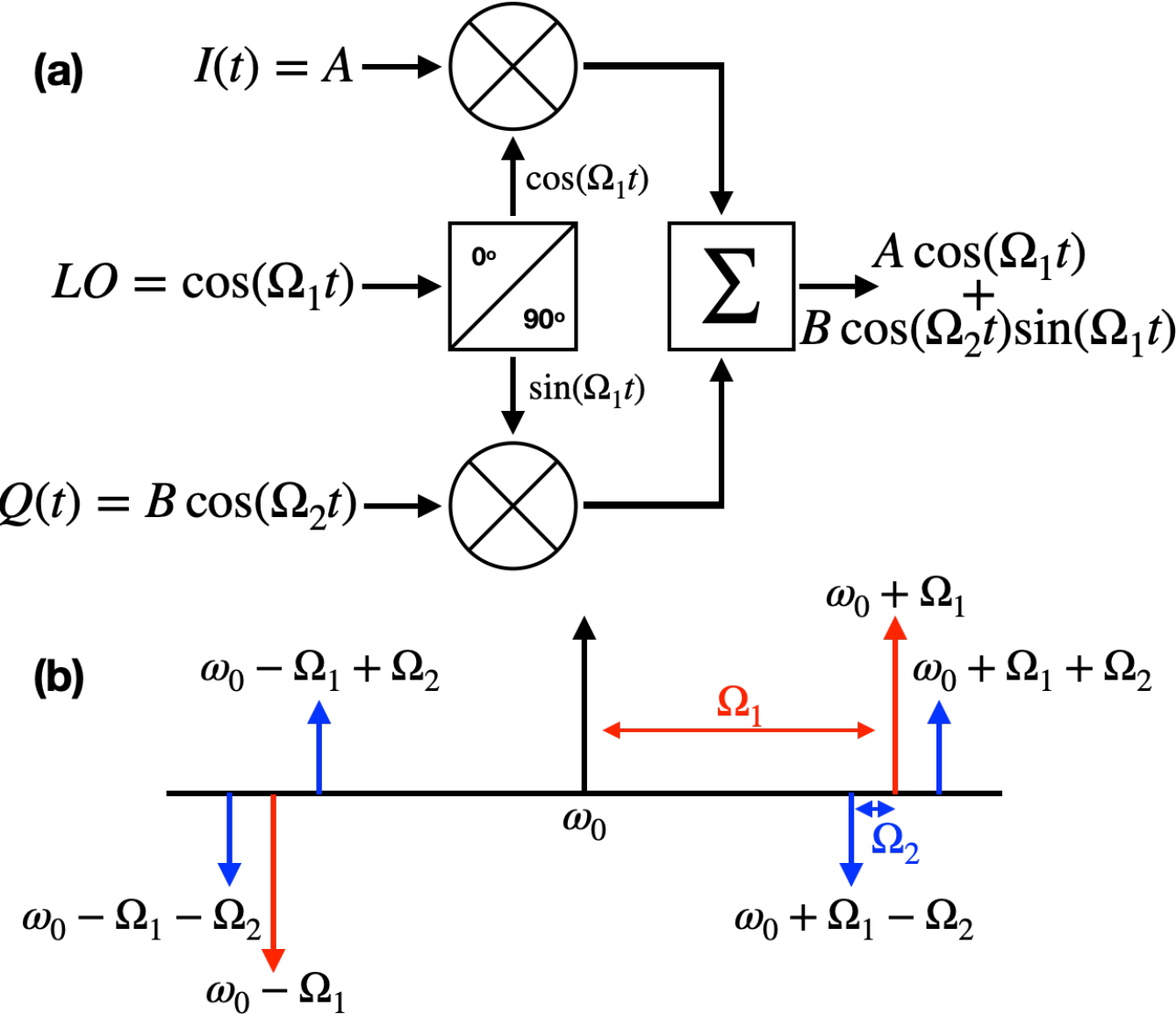}
    \caption{(a) A schematic showing the generation of sidebands using an IQ modulator. An RF signal at $\Omega_1$ fed into the LO port is split into the in-phase and quadrature components. The quadrature part is modulated with an RF signal at $\Omega_2$ and combined with the in-phase component. (b) The frequency sidebands created on an input laser frequency at $\omega_{0}$ by the IQ modulated EOM up to $\mathcal{O}\left(J_{1}\right)$ in Eq.~\ref{eqn:5}.}
    \label{fig:IQmod}
\end{figure}

An IQ modulator splits the input carrier radio frequency signal into its In-phase (I) and Quadrature (Q) components. With appropriate choices of inputs into the I and Q ports it can approximate frequency modulation. In our case, to generate the necessary frequency sidebands, a carrier with frequency $\Omega_1$ is fed into the local oscillator (LO) port of the IQ modulator. A constant DC voltage is fed into the I port while a modulation signal with frequency $\Omega_2$ is fed into the Q port. The resultant combination of the in-phase and quadrature signals is then applied to the EOM. The principle of operation of an IQ modulator is shown in Figure~\ref{fig:IQmod}a.

The electric field of a laser beam with frequency $\omega_{0}$ and amplitude $E_{0}$ can be written as
\begin{equation}
    E(t) = E_{0}e^{i\omega_{0} t}
\end{equation}
For the ESB offset lock scheme, the IQ modulated signal $A\cos(\Omega_1t)+B\cos(\Omega_2t)\sin(\Omega_1t)$ is amplified and applied to the EOM. The modulated electric field of the laser beam is then given by
\begin{equation}
\label{eqn:2}
    E(t) = E_{0}e^{i\{\omega_{0} t + A\cos(\Omega_1t)+B\cos(\Omega_2t)\sin(\Omega_1t)\}}
\end{equation}
where, $\Omega_{1}$ and $\Omega_{2}$ are the respective modulations into the LO and the Q ports of the  IQ modulator with associated modulation depths $A$ and $B$ respectively. Expanding the above expression in terms of Bessel functions, with terms up to $\mathcal{O}\left(J^{2}_{1}\right)$, enables us to see the associated sidebands generated. 

\begin{align}
    E(t) \approx E_{0}&\left\{J_{0}(A)J^{2}_{0}(B/2)e^{i\omega_{0} t} \right. \nonumber\\
    &+ \left.iJ^{2}_{0}(B/2)J_{1}(A)\left[e^{i(\omega_{0}+\Omega_{1})t} + e^{i(\omega_{0}-\Omega_{1})t}\right] \right. \tag{3a}\\
    &+ \left.J_{0}(A)J^{2}_{1}(B/2)\left[e^{i(\omega_{0}+2\Omega_{1})t}+e^{i(\omega_{0}-2\Omega_{1})t}-e^{i(\omega_{0}+2\Omega_{2})t}-e^{i(\omega_{0}-2\Omega_{2})t}\right] \right. \tag{3b}\\
    &+ \left.J_{0}(B/2)J_{1}(B/2)\left(J_{0}(A)-2J_{1}(A)\sin(\Omega_{1} t)\right)\times \right. \nonumber\\
    &\left. \left[e^{i(\omega_{0}+\Omega_{1}-\Omega_{2})t}+e^{i(\omega_{0}+\Omega_{1}+\Omega_{2})t}-e^{i(\omega_{0}-\Omega_{1}+\Omega_{2})t}-e^{i(\omega_{0}-\Omega_{1}-\Omega_{2})t}\right] \right\} \label{eq:IQEfield} \tag{3c}
\end{align}

where $J_{n}$ is the $n$-th order Bessel function of the first kind. Up to terms linear in $J_{1}$, we see the six sidebands generated as shown in Figure~\ref{fig:IQmod}b. {This resultant complex modulation structure can be compared to the ``dual sideband'' (DSB) technique presented in Ref.\cite{Thorpe08}. However, unlike in the DSB scheme, the IQ modulation does not produce sidebands about the carrier at the modulation frequency $\Omega_{2}$, and the subsequent demodulation does not generate an error signal when the carrier is resonant with one of the cavity modes. For the DSB lock, this can be particularly detrimental for cavities with large numbers of transverse modes, and therefore requires that the carrier is sufficiently suppressed to mitigate this effect. However, the IQ modulation scheme is not susceptible to this effect.} 

We also note in Eq.~\ref{eq:IQEfield}, that in terms up to $\mathcal{O}\left(J^{2}_{1}\right)$ we observe amplitude modulations at $\Omega_{1}$. This is a consequence of using an IQ modulator for phase-modulations, there is always some residual amplitude modulation (RAM). {The EOM is thermally stabilized and light polarization is adjusted to minimize RAM on the EOM output. We expect that active stabilization  could further reduce RAM \cite{Zhang14}.} For the present purposes, we focus our attention on the modulation at $\omega_{0}-\Omega_{1}$ and the two sidebands generated at $\omega_{0}-\Omega_{1}-\Omega_{2}$ and $\omega_{0}-\Omega_{1}+\Omega_{2}$. This is similar to the sidebands generated for the PDH scheme. The modulation at $\Omega_{2}$ generates the error signal, while the modulation at $\Omega_{1}$ provides the tunability of the lock. Compared to a conventional phase-modulated PDH frequency stabilization scheme with a modulation at frequency $\Omega_{1}$ with a modulation-depth $A$, the relative size of the error signal (up to $\mathcal{O}\left(J^{2}_{1}\right)$) is given by

\begin{equation}
    D_{\mathrm{ESB}}/D_{\mathrm{PDH}} = J^{3}_{0}(B/2)J_{1}(B/2)\left[1 - \frac{2J_{1}(A)}{J_{0}(A)}\sin(\Omega_{1}t)\right] \tag{4}
\end{equation}

\begin{figure}[ht]
    \centering
    \includegraphics[width=0.9\textwidth]{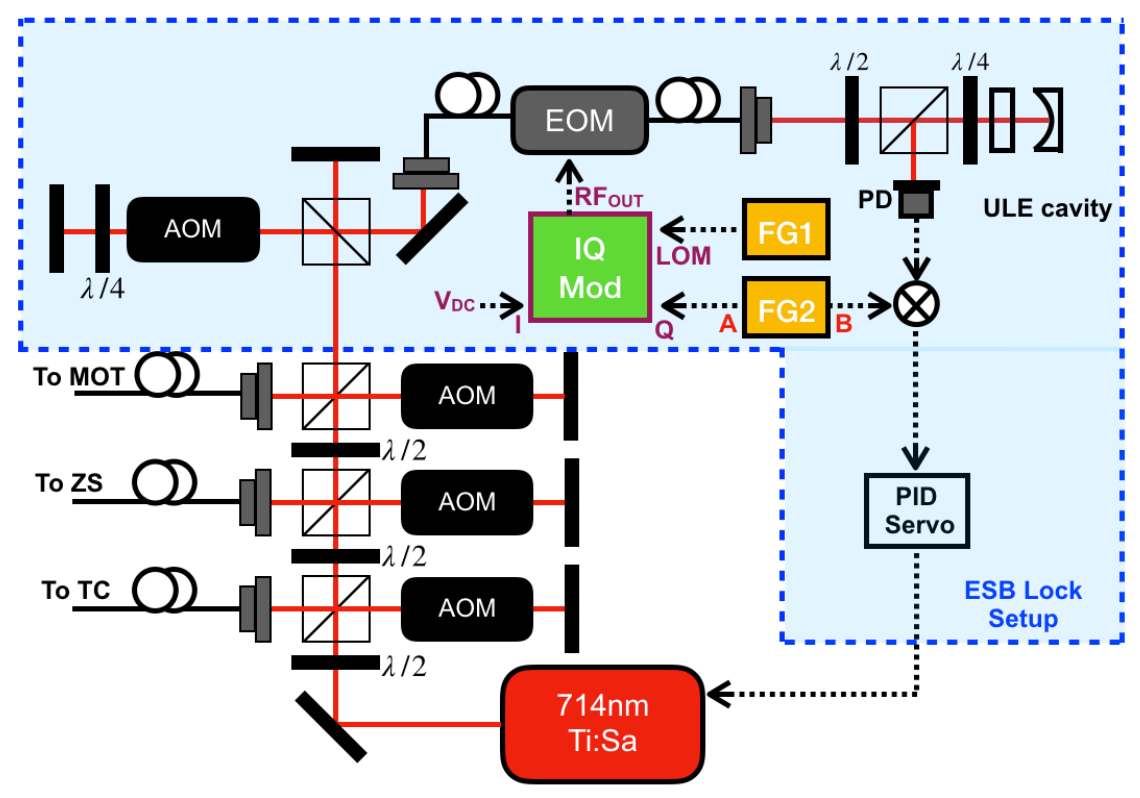}
    \caption{ESB Lock setup for our 714 nm Ti:Sa ring cavity laser. The laser output power is divided amongst the transverse cooling (TC), Zeeman slower (ZS), MOT, and ESB lock setup. The EOM is driven with the output of the IQ modulator that modulates the laser beam incident on the ULE reference cavity. {Two separate function generators (FG1 and FG2) provide the signals for the local oscillator input (LOM) and the quadrature modulation input (Q) respectively. The light reflected from the optical cavity is collected on the photo detector (PD). The PD signal is demodulated and sent to a PID controller for frequency stabilization.} The AOM before the EOM input fiber allows us to avoid cavity mode collisions as described in Section 2.}
    \label{fig:ESBsetup}
\end{figure}

By tuning the EOM offset frequency $\Omega_{1}$ we change the offset of the laser frequency from the resonant optical cavity mode and therefore tune the laser frequency. The collision of the different sidebands as one tunes $\Omega_{1}$ poses a potential issue for broadband applications. In the above case, when $\Omega_{1}$ is a multiple of $\Delta\nu_{\mathrm{FSR}}/2$, where  $\Delta\nu_{\mathrm{FSR}}$ is the free spectral range of the optical cavity, we notice the collision of the sidebands with opposite phases at $\omega_{0}+\Omega_{1}$ and $\omega_{0}-\Omega_{1}$. This can significantly affect the error signal and therefore the lock performance. However, we find that adding an AOM in the pathway helps optimize the tuning frequency $\Omega_{1}$ and, as a result, prevents any lock degradation (See Fig~\ref{fig:ESBsetup}). Incidentally, this effect can also be used to conveniently and accurately measure $\Delta\nu_{\mathrm{FSR}}$ of the cavity. One can tune the laser so that it scans a range near the center of two $\text{TEM}_{00}$ modes, then adjust $\Omega_{1}$ until the error signal of the lower frequency mode precisely cancels that of the higher order mode. The free spectral range is then twice $\Omega_{1}$. Greater relative accuracy can be obtained by comparing non-adjacent modes, and measuring 4 or 5 times $\Delta\nu_{\mathrm{FSR}}$. 

{It is also possible to reduce mode collisions and increase error signal size using serrodyne modulation for $\Omega_{1}$ \cite{Kohlhaas:12}. This technique uses a non-linear transmission line (NLTL) to create a "sawtooth" waveform for $\Omega_{1}$ rather than a sinusoidal function. Modulating with a function of this form can result in a single sideband at $\omega_{0}-\Omega_{1}$ \textit{or} $\omega_{0}+\Omega_{1}$ rather than two sidebands at $\omega_{0}\pm\Omega_{1}$.  We chose not to adopt this technique due to the added complexity of adding the NLTL.  Furthermore, since the sawtooth pattern depends on the presence of several harmonics of $\Omega_{1}$, the advantage of serrodyne modulation would only be possible over a fraction of the bandwidth of the IQ modulator with the potential for unpredictable behavior when harmonics present in the sawtooth waveform exceed the bandwidth of the IQ modulator.  This is especially a concern for isotope shift spectroscopy where $\Omega_{1}$ can change by a large fraction of the IQ modulator bandwidth}

\section{MOT cutoff and isotope shift measurements in $^{225}\text{Ra}$ and $^{226}\text{Ra}$}

The following section describes the implementation of an ESB offset lock in our 714 nm laser system and its application to {isotope shift} spectroscopy in \raI\, and \raII.
  
\subsection{ESB Offset Lock Setup}

The experimental setup used for the ESB offset lock of our 714 nm laser and {for radium spectroscopy} is shown in Fig~\ref{fig:ESBsetup}. We generate $\thicksim$1.3 W of 714 nm light from a Ti:Sapphire ring cavity laser (Sirah, Matisse) pumped by a diode-pumped solid state (DPSS) laser (Lighthouse Photonics, Sprout). Most of the light is sent to our laser slowing and trapping setup for radium. A small sample of $\thicksim$4 mW is sent to a fiber-coupled EOM (EOSpace). The EOM provides phase modulation to the laser beam which is sent to {a temperature-stabilized, ultra low expansion glass (ULE) optical cavity ($\Delta\nu_{\mathrm{FSR}}=1.5$ GHz, finesse $\mathcal{F}=2000\pm300$ at 714 nm) for frequency stabilization.}  

\begin{figure}[ht]
    \centering
    \includegraphics[width=0.4\textwidth]{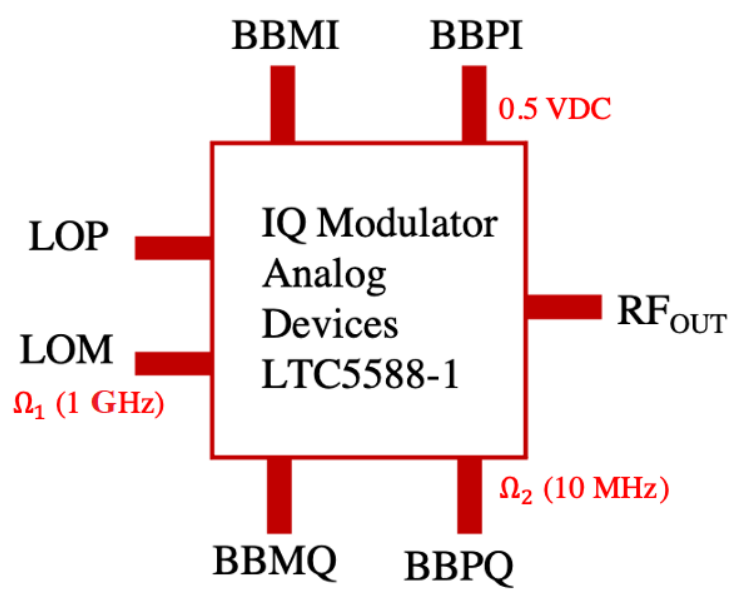}
    \caption{IQ modulator (Analog Devices, LTC5588-1) with the relevant ports and inputs are shown.  The $\Omega_{1}$ modulation (1GHz, 0.5Vpp) is fed into the LOM (negative LO input) port of the IQ modulator and the LOP (positive LO input) port is terminated at $50\,\Omega$. A constant 0.5 VDC is applied to the BBPI input port of the I-channel of the IQ modulator, and the BBMI port is terminated at $50\,\Omega$. The $\Omega_{2}$ modulation (FG2) (10 MHz, 0.5 Vpp) is fed into the BBPQ port of the Q-channel of the IQ modulator, and the BBMQ port is DC biased so as to maintain a common mode voltage of 0.5 VDC. The resultant modulation through the $\text{RF}_{\text{OUT}}$ port is sent to the EOM.}
    \label{fig:IQmodpinout}
\end{figure}
\begin{figure}[ht]
    \centering
    \includegraphics[width= 0.6\textwidth]{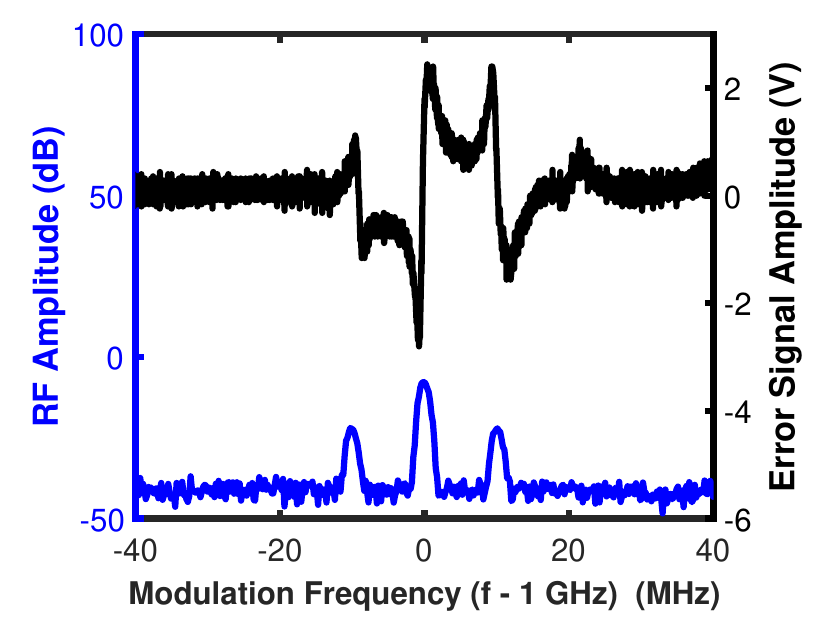}
    \caption{Blue Trace: Output of the IQ modulator showing the  modulation sideband at $\Omega_1$ with the two sidebands at $\Omega_1\pm\Omega_2$. $\Omega_1$ is set to 1 GHz and $\Omega_2$ to 10 MHz. Black Trace: The error signal generated after demodulating the reflected optical signal at $\Omega_2$.}
    \label{fig:714errorsignal}
\end{figure}

The modulation RF signal is generated by an IQ modulator (Analog Devices, LTC5588-1). A schematic of the IQ modulator is shown in Figure~\ref{fig:IQmodpinout} along with the relevant ports and inputs.
The blue trace in Figure~\ref{fig:714errorsignal} shows a typical output from the IQ modulator. Here we see the offset sideband at 1 GHz ($\Omega_1$) and the corresponding modulation sidebands that are offset $\pm10$ MHz ($\Omega_2$) from $\Omega_1$.  The reflected light from the cavity is collected on a fast photodiode (PD). The signal is sent to a mixer and demodulated at 10 MHz to create an error signal as shown by the black trace in Figure~\ref{fig:714errorsignal}.

The error signal is low-pass filtered and sent to a PID controller inside the Ti:Sapphire laser control box, which feeds back on the fast etalon in the ring cavity to lock the laser frequency to a TEM$_{00}$ mode of the ULE cavity. {Using the known finesse and free spectral range of the the ULE cavity, we estimate the laser linewidth to be $\thicksim$70 kHz by comparing the RMS value of the error signal when locked to its peak-to-peak value as we scan the laser across the cavity resonance.}
 
\subsection{MOT cutoff measurement setup}

\begin{figure}[ht]
    \centering
    \includegraphics[width= 0.5\textwidth]{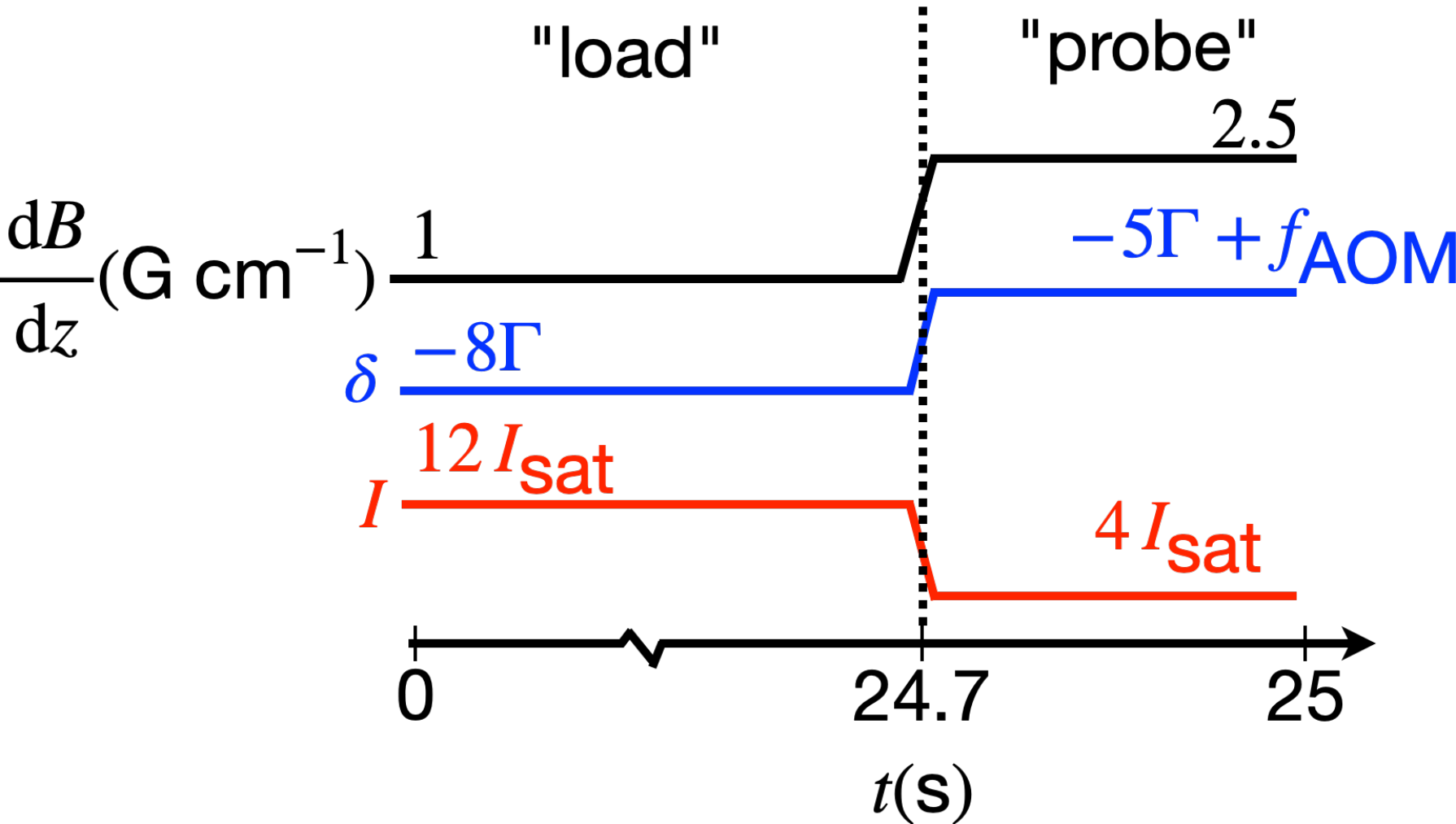}
    \caption{MOT phases during an experimental cycle. We begin with the MOT in the ``load'' phase ($\frac{dB}{dz}= 1\,\text{Gcm}^{-1}$, $\delta = -8\Gamma$, $I = 12I_{\text{sat}}$) for 24.7 s, and switch to the probe phase ($\frac{dB}{dz} = 2.5$ G cm${}^{-1}$, $\delta = -5\Gamma + f_{\text{AOM}}$, $I = 4I_{\text{sat}}$) for the remaining 0.3 s. During the probe phase we scan the RF modulation frequency for the MOT AOM ($f_{\text{AOM}}$) and measure the atomic fluorescence.}
    \label{fig:expsequence}
\end{figure}

Most of the 714 nm laser power is sent to the transverse cooling (TC) chamber to collimate the atomic beam coming out of the oven, followed by the Zeeman slower (ZS) to slow the longitudinal velocity of the atomic beam below the MOT capture velocity, and finally to the MOT to trap the radium atoms. More details about the experimental design and the slowing and trapping of radium can be found in Ref.~\cite{mike16}. {As shown in Fig.~\ref{fig:expsequence}, our experimental sequence consists of two ``phases'' and lasts for 25 s.  The sequence begins with a ``load'' phase where the MOT laser beam intensity, $I=12I_{\mathrm{sat}}$, B-field gradient, $\frac{dB}{dz}=1$ G cm${}^{-1}$, and detuning, $\delta = -8\Gamma$, are set to optimize atom capture into the MOT.  Here, $I_{\mathrm{sat}}=140$ $\mu$Wcm${}^{-2}$ is the saturation intensity and $\Gamma = 2\pi \times 380$ kHz is the natural linewidth of the \singletSzero\ to \tripletPone\ transition.  After 24.7 s we switch to the ``probe'' phase for 0.3 s during which the intensity, gradient, and detuning are set to $I=4I_{\mathrm{sat}}$, $\frac{dB}{dz}=2.5$ G cm${}^{-1}$, and $\delta = \delta_\text{probe}$, respectively.  We adjust $\delta_\text{probe}$ by changing the frequency applied to the MOT AOM ($f_{\text{AOM}}$).  We quickly switch $f_{\text{AOM}}$ between loading and a probe phase values using two separate RF generators connected to the AOM through an RF switch.  During the probe phase, we measure the MOT fluorescence on an EMCCD camera (Andor, Luca) for different values of $f_{\text{AOM}}$.}  

For $^{225}$Ra, we further offset the laser frequency from the cavity reference mode by 2629.95 MHz. This is achieved by increasing the offset frequency $\Omega_1$ by the same amount. This highlights the broadband tunability of this implementation of the ESB lock. Locking to the same cavity resonance, we simply increase the laser frequency offset to probe transitions that are several GHz apart. Using the same cavity resonance eliminates any potential systematic or statistical frequency uncertainties introduced by referencing the laser to different cavity resonances. 

\subsection{Data analysis}

We measure the MOT fluorescence signal for $^{225}\text{Ra}$ and $^{226}\text{Ra}$ during the probe phase. For each isotope, we collect 5 background images before and after the frequency scan. We scan the MOT probe frequency in steps of 20 kHz to 50 kHz by scanning the RF drive frequency of the MOT AOM {($f_{\text{AOM}}$)} during the probe phase. At each frequency step during the scan, we acquire an atom fluorescence image. The atom images are background subtracted to extract the atom fluorescence count. 

\begin{figure}[ht]
    \centering
    \begin{subfigure}[b]{0.48\textwidth}
        \caption{$^{226}$Ra}
        \includegraphics[width=\textwidth]{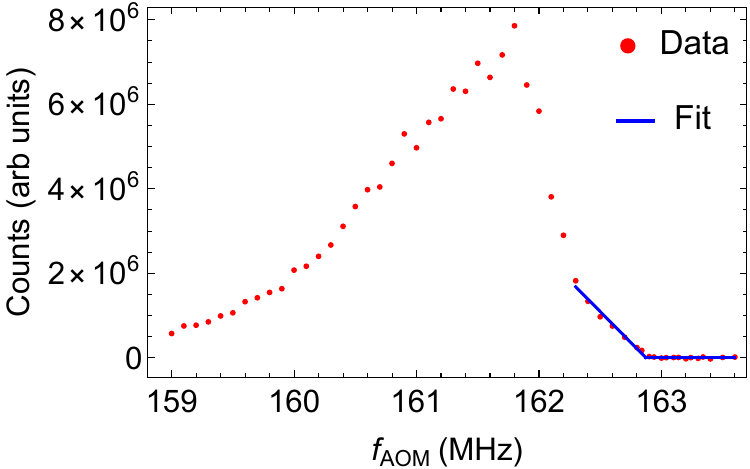}
        \label{fig:mot226}
    \end{subfigure}
    \hfill
    \begin{subfigure}[b]{0.49\textwidth}
        \caption{$^{225}$Ra}
        \includegraphics[width=\textwidth]{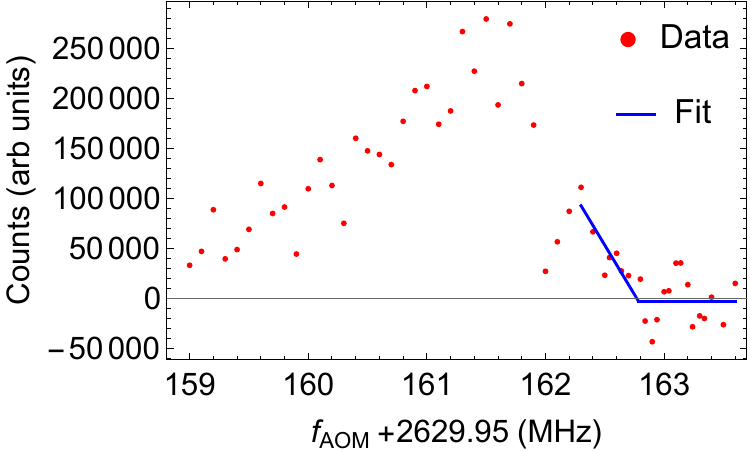}
        \label{fig:mot225}
    \end{subfigure}
    \caption{Measured MOT fluorescence counts versus AOM frequency {($f_{\text{AOM}}$)} for (a) $^{226}$Ra and (b) $^{225}$Ra. The laser frequency for $^{225}$Ra is further offset from the cavity resonance relative to $^{226}$Ra by increasing the EOM offset frequency $\Omega_1$ by 2629.95 MHz. The difference in the fluorescence counts of the two isotopes is primarily due to the different number of atoms loaded into the oven.}
    \label{fig:motcutoff}
\end{figure}

The MOT cutoff frequency is defined as the laser frequency at which the MOT fluorescence vanishes as the frequency is increased. As depicted in Figure~\ref{fig:mot226}, above a certain frequency, the MOT fluorescence counts drop below the background level.  
{For previous measurements of the \raII\ EDM \cite{mike16,Parker2015}, a daily determination of the MOT cutoff frequency in \raI\ was used to adjust the cavity offset frequency to compensate for the frequency drift of our ULE cavity (typically <10 kHz per day).  This was necessary to achieve optimal cooling from the MOT into an optical dipole trap and transfer between two different optical dipole traps (optical dipole traps are not used in the current work), both of which are sensitive to frequency deviations below 100 kHz \cite{Parker2012}.  The success of this procedure gives us confidence that our MOT cutoff frequency measurement is repeatable at this level. Moreover, the additional signal afforded by performing spectroscopy within an active MOT is essential to detecting the small number of \raII\ atoms in our current apparatus.  It should be noted that the MOT cutoff frequency is not necessarily the transition line center, however, we  consider various possible systematic effects below and do not find any significant effect at the current level of precision.  }

{The MOT cutoff frequency measurements also benefit from several advantageous properties of the radium transitions we interrogate.  First, the Land\'{e} g-factor for the \tripletPone\ $F=3/2$ state in \raII\ is $g_{F} = (2/3)g_{J}$, neglecting corrections $\mathcal{O}(\frac{\mu_N}{\mu_B})$. Here, $g_{J}$ is the corresponding g-factor for \raI, $\mu_N$ is the nuclear magneton, and $\mu_B$ is the Bohr magneton.  The MOT primarily excites atoms into the, $m_F=\pm 3/2$ states due to the combination of magnetic field gradient, laser polarization and optical pumping.  The Zeeman shift in these states is proportional to $g_{F}m_{F} = (2/3)g_{J}(3/2) = g_{J}$ and, as a result, we expect similar field-dependent shifts in both isotopes. Furthermore, the isotope masses only differ by 0.4\%. The shape of the fluorescence signal is modified at the $\sim$MHz level (see Figure~\ref{fig:motcutoff} below) by the field gradient and scattering force in the MOT.  Thus, we expect mass-dependent frequency shifts due to differential acceleration forces in the MOT to be less than 10 kHz.  Finally, the effect of the $F=1/2$ hyperfine state present in \raII\ (and absent in \raI) is found to be negligible for our measurements by considering the scattering rate of the MOT laser for the  $F=1/2$ state when it is tuned near the $F=3/2$ state. The two states are separated by 15 GHz so the detuning from the $F=1/2$ state for this work is given by $\delta_{F=1/2} =2\pi\times15$ GHz $= 40,000\Gamma$.  Even at the largest MOT probe power used in this work (25.8 mW) this corresponds to a scattering rate of 0.003 per second making this transition completely negligible for affecting our fluorescence signal or MOT dynamics.}

To determine the MOT cutoff frequency, we assume the fluorescence signal is linear in the region of the cutoff and fit to the following function the segment of the MOT fluorescence curve containing several data points just before the fluorescence curve flattens: 

\begin{equation}
    {y(f_{\text{AOM}}) = \text{Max}\left[b(f_{\text{AOM}} - a_{1}), b(a_{2} - a_{1}) \right]} \tag{5}
    \label{eqn:5}
\end{equation}

This fit function is the maximum between a line with a slope, $b$, and x-intercept, $a_{1}$, and a line with a constant y value given by $b(a_2-a_1)$. The MOT cutoff frequency, represented by $a_2$, is defined as the frequency at which these two lines intersect. 
The linear approximation is valid for only those data points near the cutoff frequency.  We therefore studied the sensitivity of the fitted value of MOT cutoff frequency on the number of data points used for the fit, which is discussed in the next section.

\subsection{Results and discussion}

{The MOT fluorescence data we use to determine the isotope shift between \raI\ and \raII\ are shown in Fig~\ref{fig:motcutoff}.  Both data sets are acquired on the same day to eliminate additional uncertainty due to ULE cavity drift.  We also use the same probe power ($\sim16$ mW) and MOT B-field gradient (2.5 G cm${}^{-1}$) for both isotopes. From a fit to Equation \ref{eqn:5} we determine the MOT cutoff frequency for $^{226}$Ra (with statistical uncertainties) to be $\nu_{226} = 162.87(1)$ MHz and for $^{225}$Ra to be $\nu_{225}+2629.95 \,\text{MHz} = 162.79(10)$ MHz. The frequency difference for the MOT transition between the two isotopes is therefore determined to be 2630.03 MHz with a statistical uncertainty of $\sigma_{\Delta\nu,\mathrm{stat}} = 0.10\,\text{MHz}$.}

\begin{figure}[ht]
    \centering
    \includegraphics[width= 0.6\textwidth]{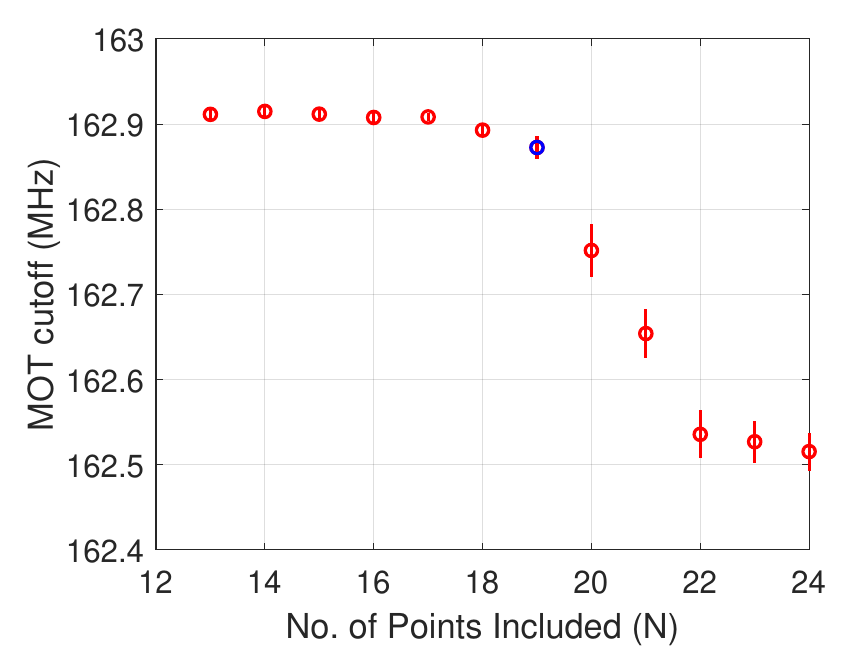}
    \caption{MOT cutoff frequency {$a_2$} for $^{226}$Ra (probe power at $\sim16$ mW) as a function of the number of data points included in the fit.  We include the data points {corresponding to the 12 largest $f_{\text{AOM}}$ values and additional points moving down in $f_{\text{AOM}}$ up to the point of peak fluorescence signal}. We fit the data up to the point before the MOT cutoff frequency value diverges by more than a standard deviation. In this case, for N greater than 19, the value of $a_2$ diverges and so only up to the N = 19 data point (blue data point) are included in the fit.}
    \label{fig:motcutoff226_vs_pts}
\end{figure}

{To select the number of data points to include in our fit, we study the relationship between the number of data points included in the fit to Eqn.~\ref{eqn:5} and the fitted value for MOT cutoff frequency, $a_2$. For example, we fit the $^{226}$Ra dataset shown in Fig.~\ref{fig:mot226}, including the 12 data points corresponding to the highest AOM frequencies as well as a varying number of points moving down in AOM frequency to the point of peak fluorescence signal and we observe how the fitted value of $a_2$ changes. We expect the fit values for $a_2$ to converge to the true value when the lowest AOM frequency point is near the MOT cutoff frequency so the linear approximation remains valid, and to diverge as we include more points and this approximation breaks down. As shown in Figure~\ref{fig:motcutoff226_vs_pts}, the MOT cutoff frequency fit values diverge by more than one standard deviation after more than 19 data points are included in the fit. We therefore fit for the MOT cutoff frequency $a_2$ using these 19 data points for the above dataset. The data and the resultant fit are shown in Figure~\ref{fig:mot226}. This method for determining the cutoff frequency is used for all the MOT fluorescence data although only the resulting MOT cutoff frequencies from the data included in Figure~\ref{fig:motcutoff} are used to determine the frequency shift between the \raI\ (\singletSzero\ to \tripletPone) and \raII\ (\singletSzero\ $F=1/2$ to \tripletPone\ $F=3/2$) MOT transitions.}

\begin{figure}[ht]
    \includegraphics[width=\textwidth]{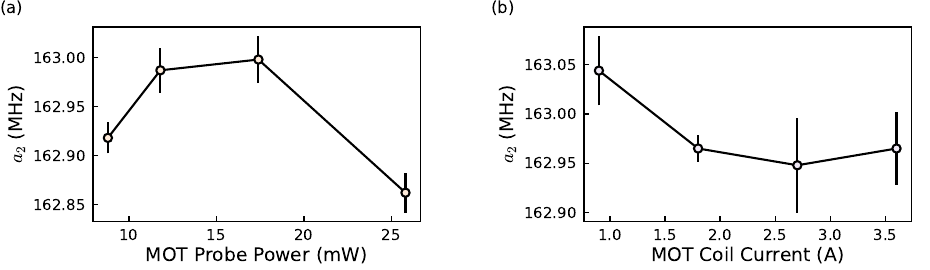}
    \caption{{The MOT cutoff frequency ($a_{2}$) for \raI\, is plotted for different values of (a) MOT laser power and (b) MOT B-field coil current that are applied during the probe phase of the experimental sequence. We do not observe a significant effect for either parameter, so we use the largest observed variation in MOT cutoff frequency as each parameter is varied to place an upper limit on the size of these effects.}}
    \label{fig:motcutoffsystematics}
\end{figure}

{The limited availability and short half-life of \raII\ ($\tau_{1/2}=14.9$ days) make broad, exploratory, systematic investigations impractical when using this isotope.  Thus, we use \raI\ to explore possible systematic effects and determine if they exist at a detectable level. We consider systematic shifts of the MOT cutoff frequency that depend on MOT laser intensity and applied B-field gradient during the probe phase.  To investigate laser intensity, we keep the probe MOT gradient B-field constant and measure MOT fluorescence spectra using different MOT laser powers. The dependence of the cutoff frequency on the probe MOT power for $^{226}$Ra is shown in Figure~\ref{fig:motcutoffsystematics}a. We consider total powers between 8.8 mW and 25.8 mW corresponding to single beam peak intensities between $2I_{\mathrm{sat}}$ and $6I_{\mathrm{sat}}$. The data do not show a significant trend proportional to laser power so we estimate an upper limit for this effect by taking the difference between the maximum and minimum MOT cutoff values from the four different MOT powers studied (including error bars). This conservatively limits any shift of the MOT cutoff freqeuncy that depends on probe power to below 0.18 MHz.}

{We next investigate the effect of the MOT B-field gradient on the observed cutoff frequency by measuring MOT fluorescence spectra using different currents applied to the anti-Helmholtz MOT B-field coils during the probe phase. We vary the current between 0.9 and 3.6 A, corresponding to gradients between 1.1 and 4.5 G cm${}^{-1}$.  The plot of the MOT cutoff frequency fit values for the different MOT B-field coil currents are shown in Figure~\ref{fig:motcutoffsystematics}b. Again, we do not observe a significant trend so we estimate a conservative upper limit for this effect using the same procedure as for probe MOT power and we get an identical upper limit, 0.18 MHz. In future works, greater precision in systematic evaluations could be achieved by testing greater powers or field gradients, obtaining more fluorescence data, or using alternate techniques for performing spectroscopy with MOT-trapped atoms \cite{Goncharov_2014}.  For the current work, we add in quadrature, the upper limit values determined for both effects to determine the total systematic uncertainty, $\sigma_{\Delta\nu,\text{sys}} = 0.25\,\text{MHz}$. Adding this in quadrature to the statistical uncertainty of $\sigma_{\Delta\nu,\text{stat}} = 0.10\,\text{MHz}$ gives a total uncertainty of our MOT cutoff frequency difference between the two isotopes of $\sigma_{\Delta\nu} = 0.27\,\text{MHz}$. The frequency difference for the MOT transitions between \raI (\singletSzero $\rightarrow$ \tripletPone) and \raII (\singletSzero [F=1/2] $\rightarrow$ \tripletPone [F=3/2]) is therefore determined to be}

\begin{align}
    \Delta \nu &= \nu_{226} - \nu_{225} \nonumber \\
                &= 2630.0 \pm 0.3 \, \text{MHz} \tag{6}
\end{align}
    
This is consistent with the value calculated from previously available spectroscopic data for radium~\cite{Wendt87} of $2629.0\pm8.6$ MHz and is a factor of 29 more precise. Additionally, along with the known hyperfine splitting between the $F=1/2$ and the $F=3/2$ levels of \tripletPone \cite{Scielzo06}, we use our measurement of the difference in the MOT transitions between \raI\, and \raII\, to calculate the center-of-gravity isotope shift, $\Delta \nu_{\text{iso}}$, between these two isotopes for the \singletSzero to \tripletPone transition:
\begin{align}
    \Delta \nu_{\text{iso}} &= \frac{1}{3}\Delta \nu_{\text{hfs}} - \Delta \nu \nonumber \\
    &= \frac{14691\pm6.5}{3} - 2630\pm0.3  \tag{7}\\
    &= 2267.0\pm2.2\, \text{MHz}\nonumber
\end{align}
{where the resultant uncertainty is predominantly due to the 6.5 MHz uncertainty on the known hyperfine splitting between the \tripletPone $F=1/2$ and $F=3/2$ levels in \raII. This is a factor of 8 more precise than the isotope shift of $2268\pm17$ MHz that can be determined from Ref.~\cite{Wendt87}. It should be noted that, in Ref.~\cite{Wendt87}, the isotope shifts for the \singletSzero to \tripletPone transition were measured for a range of radium isotopes with respect to $^{214}$Ra. Therefore it is possible that the uncertainty of 17 MHz propagated onto this isotope shift between \raI\, and \raII\, is overestimated. Since, to the best of our knowledge, no direct measurements of this isotope shift is reported, we take the above calculated value and its propagated uncertainty to be the currently best available measurement.  Future iterations of our experiment with larger atom number and IQ modulator bandwidth could greatly reduce the uncertainty in isotope shift using our ESB offset locking technique, at which point second order hyperfine shifts may become relevant \cite{Beloy08}}.


\section{Conclusion}

We have demonstrated the implementation of a broadband, tunable ESB offset lock using an IQ modulation scheme. 
{We employ this laser stabilization scheme with a 714 nm laser to determine the frequency shift between MOT transitions in \raI\ and \raII. The precision of our measurement greatly improves upon previous measurements because our ESB lock allows us to stabilize our laser to a single resonance of a ULE optical cavity with negligible drift on the timescale of our measurements while tuning our laser by over 2 GHz to address both transitions. While the improved measurement precision we demonstrate for a single transition and a single pair of isotopes cannot dramatically improve our understanding of radium nuclei, future work leveraging our stabilization scheme could provide greatly improved measurement precision across a broad array of transitions, isotopes, and even other elements.  Such precision could uncover perturbations caused by higher order nuclear moments \cite{Beloy08} that illuminate the structure of nuclei with unprecedented precision.}

The ESB locking technique we use is also broadly applicable to myriad other physical systems that would benefit from improved precision in  differential frequency measurements and could be expanded to frequency differences beyond 6 GHz through the use of higher bandwidth IQ modulators. We expect this technique will not only significantly improve precision in isotope shift spectroscopy in particular, but could have broader impacts in laser spectroscopy more generally.

\begin{backmatter}
\bmsection{Funding}
This work is supported by the U.S. DOE, Office of Science, Office of Nuclear Physics under contracts DE-AC02-06CH11357 and DE-SC0019455, and by Michigan State University.

\bmsection{Acknowledgments}
\raII\ used in this research was supplied by the U.S. Department of Energy Isotope Program, managed by the Office of Isotope R\&D and Production.

\bmsection{Disclosures}
The authors declare no conflicts of interest

\bmsection{Data Availability} Data underlying the results presented in this paper are not publicly available at this time but may be obtained from the authors upon reasonable request.

\end{backmatter}


\bibliography{sample}






\end{document}